\documentclass[amsmath,amssymb,reprint,bibnotes]{revtex4-2}
\usepackage{graphicx}
\usepackage{bm}
\usepackage{amsmath} 
\usepackage{graphics}
\usepackage[dvipsnames]{xcolor}
\usepackage[mathlines]{lineno}
\usepackage{braket}
\usepackage[compact]{titlesec}
\begin{document}
\title{Voltage-controlled extraordinary optical transmission in the visible regime}
\author{Hira Asif$^{\bf (1)}$}
\author{Alpan Bek$^{\bf (2)}$}
\author{Mehmet Emre Tasgin$^{\bf (3)}$}\email{metasgin@hacettepe.edu.tr}
\author{Ramazan Sahin$^{\bf (1)}$}\email{rsahin@itu.edu.tr}

\affiliation{${\bf (1)}$ {Department of Physics, Akdeniz University, 07058 Antalya, Turkey}}
\affiliation{${\bf (2)}$ {Department of Physics, Middle East Technical University, 06800 Ankara, Turkey}}
\affiliation{${\bf (3)}$ {Institute of Nuclear Sciences, Hacettepe University, 06800 Ankara, Turkey}}

\date{\today}

\begin{abstract}
Control of components in integrated photonic circuits is crucial in achieving programmable devices. Operation bandwidth of a plasmonic device cannot be generally tuned once it is manufactured, especially in the visible regime. Here, we demonstrate the electrical control of such a device for extraordinary optical transmission~(EOT) in the visible regime. (i) Operation frequency of the EOT device can be tuned via a bias voltage applied through nanowires. (ii) Or, at a given frequency, the EOT signal (normalized to the incident field) can be tuned continuously, e.g., between $10^{-4}$ and $0.4$. This corresponds to a 3-orders of magnitude modulation depth. We utilize Fano resonances induced by a quantum emitter~(QE) that is embedded into the nanoholes. The external bias-voltage tunes QE's resonance. We also discuss the lifetime extensions of surface plasmon polaritons as a response to an ultra-short optical pulse. Our proposed method provides the active electronic control of EOT signal which makes it a feasible and compact element in integrated photonic circuits, for bio-sensing, high resolution imaging, and molecular spectroscopy applications. 	
\end{abstract}

\maketitle

\section{Introduction}
Active tuning of plasmon-based nanophotonic devices has opened new avenues for on-demand coherent control of photonic integrated circuits (PIC) \cite{Fang2015, Giordani2023}. These circuits, operating on the quantum level, hold tremendous potential for various sensing, imaging, and spectroscopy applications \cite{Hallett2018}. To achieve transient switching or dynamic control over optical-frequency signals in PIC, a scalable approach is to utilize surface confined plasmonic excitations in an optically resonant nanostructure, which can efficiently funnel electromagnetic energy from continuum to nanometer scale and subsequently radiate it back to continuum with enhanced efficiencies. In this regard, one of the most ingenious phenomenon in the field of light-matter interaction is extraordinary optical transmission (EOT) through subwavelength hole \cite{Ebbesen1998}. The enhanced transmission of light is attributed to the excitation and propagation of surface plasmons polaritons (SPP) at the metal-dielectric interface. These SPPs tunnel through the hole, interfere with other resonant modes, and scatter to the far-field, resulting in transmission exceeding unity when normalized to the hole area \cite{Barnes2003}. Apart from SPP, the subwavelength size aperture also supports localized surface plasmon resonances (LSPR) and their coupling with SPP mediated by Fano resonance contribute to enhanced transmission \cite{Xiao2015,Sahin2020}. \\ 
\begin{figure}
\includegraphics[scale=0.57]{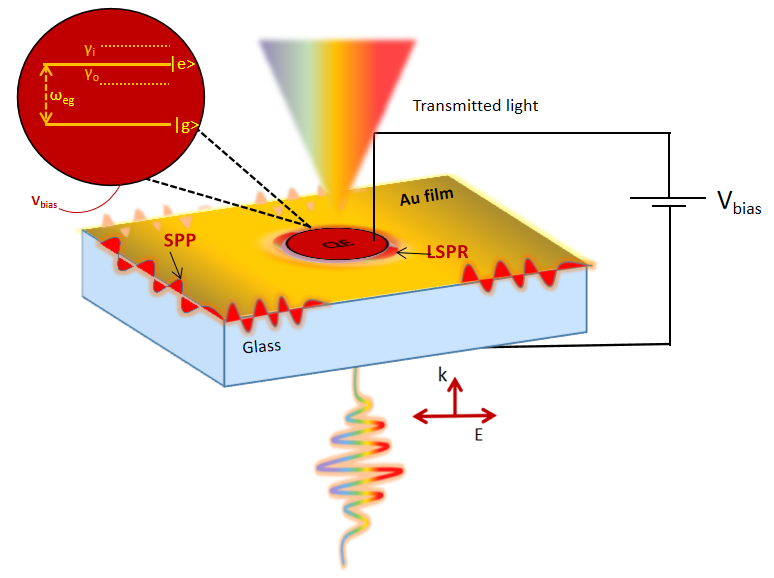}
\caption{\label{fig:1} Schematic diagram of EOT sensor. A 100 nm thick Au film, with a spherical hole (d: 200 nm), is layered on a glass substrate with Cr coating. A two-level QE is placed inside the hole and attached to external voltage bias. The excited plasmon modes near the hole edge and metal interfaces interact with voltage-tunable QE.}
\end{figure}
These characteristics SPP and LSP modes have shown their profound impact on the coherent dynamics of EOT light \cite{Liu2008}. Consequently, tailoring the spectral and temporal dynamics of these modes through external mediums enables active tuning of EOT signal. Such type of dynamical control of EOT structure not only foster chip-scale sensing but also enable wavelength sensitive photonic filters for frequency selective biosensing, high-resolution optical imaging, optical trapping and enhancement of nonlinear effects \cite{Rodrigo2016}. Recently, many studies have investigated electric, magnetic and optical control of EOT structure by employing tunable materials to actively modulate the spectral response of transmission signal from NIR to THz regime \cite{Battula2007,Shaner2007}. For instance, electrically controllable n-doped semiconductor (GaAs) material is utilized as a dielectric switch to obtain transmission in the THz regime \cite{Chen2008}. Some other studies have revealed the scope of voltage tunable graphene nanostructures for spectrally modulating the transmission signal in the THz and mid-infrared region \cite{Kim2016,Gao2021}. Despite significant enhancement in the spatial and spectral characteristics of EOT signal, these tunable mediums are limited to control ultrafast decay dynamics of plasmon resonances \cite{Koya2022}, due to their weak oscillator strength and low carrier density \cite{Kim2016}. Furthermore, most of the sensing and spectroscopy applications require tunable EOT sensors which operate in the visible spectral range with low loss, high efficiency and phase tunable capabilities \cite{Lukin2020}. To achieve such a controllable system, recently, we have proposed all-optical tuning of EOT signal in the visible regime, through ultrashort light pulse which not only facilitates enhanced efficiency but also elongates coherence time \cite{Asif2023}. However, such optical control did not account the spectral tuning of the EOT signal.\\
To acquire both spectral and temporal modulation of EOT signal, simultaneously on a single chip, here, we aim to exploit electrically tunable Quantum Emitter (QE). Quantum dots in the form of QE have emerged as a potential source for the coherent control, owing to their strong oscillator strength, high quantum efficiencies and broadband spectral tunability \cite{Lim2015,HAsif2022, Pradhan2018, Gammon1996}. 

Over the last decade, an extensive research has been done in the field of quantum plasmonics \cite{Chang2006,ZHOU2019}. Specifically, the interaction of quantum objects (such as molecules, quantum dots and nitrogen vacancies) with plasmon resonances, in the weak and strong coupling regime, has received much attention due to fluorescent enhancement, spontaneous emission control and single photon sources \cite{Hutchison2011,Akimov2007}. Whereas, in these systems ultrafast dissipation of plasmon modes and its fixed spectral response governed by geometrical parameters limits the scalability of these nanophotonic devices on a single chip \cite{Yadav2020, Wang2016, emre2022}. On the other hand, hybrid plasmon-emitter system provides a better solution for incorporating different photonic technologies into a single integrated circuit \cite{Kim2020,Vasa2018} and constitutes a scalable and on-chip integrated system for cavity-enhanced single-photon emission \cite{Bitton2022,Chakraborty2015}. In addition, an active and in-situ modulation of plasmon response is a prerequisite to achieve dynamical control of photonic systems and improve functionality for diverse quantum technologies and applications. 
	
In this paper, we introduce an electrically-tunable EOT device operating in the VIS regime. The tunability is provided over the quantum emitter~(QE) whose resonance can be controlled via an applied voltage. This can be done through tiny nanowires which makes the device compact. The device can be operated in two functionalities. (i) An already-manufactured plasmon-QE hybrid device can be operated efficiently at different incident frequencies. That is, poor EOT signal of the device at a given operation frequency can be enhanced several times by applying a proper bias voltage. The second functionality is even more useful. (ii) When the device is operated at a fixed frequency $\lambda_{\rm exc}$, the EOT signal can be electrically-tuned, e.g., between $10^{-4}$ and $0.4$ continuously (see Fig.~\ref{fig:7}). Thus, our method, employing Fano resonances, provides an indispensable EOT device which was not possible before to the best of our knowledge. We also investigate the lifetime of localized surface plasmons~(LSPs) and surface plasmon polaritons~(SPPs) as a response to an ultrafast incident pulse. By changing the transition frequency of QE through external voltage, we not only tune the spectral response of hybrid plasmon modes but also enhance the temporal bandwidth at the particular spectral shift. The coherent coupling of QE with resonant modes is optimized in two ways, the long lifetime of QE helps to enhance the oscillating time of ultrafast plasmon resonances even in a weak coupling regime and the resonant coupling manifest coherent correlation of quantized modes which induces a shift in both radiating modes independently.

\section{Analytical Approach}
We investigate the spectral dynamics of propagating (p) and localized (l) plasmon modes, excited in the EOT structure, and their interaction with QE in the light of coupled harmonic oscillator model \cite{Asif2022,Sahin2020}. For a quantized field interacting with a two-level QE, we define the Jaynes-Cummings Hamiltonian which has been widely utilized in hybrid quantum plasmonic systems \cite{Torma_2015}. The total Hamiltonian takes the form,  
\begin{eqnarray}
\hat{\mathcal{H}}_{\rm tot}=\sum_{j=p,l}\hbar\omega_{j}\hat{a}_{j}^\dagger\hat{a}_{j}+\hbar\omega_{eg}\ket{e}\bra{e}+\mathcal{M}\nonumber\\+\sum_{j=p,l}\hbar\kappa_{j}(\hat{a}_{j}^\dagger{\textdagger}\ket{g}\bra{e}+\hat{a}_{j}\ket{e}\bra{g})+\hbar f(\hat{a}_{p}^\dagger\hat{a}_{l}+\hat{a}_{l}^\dagger\hat{a}_{p})
\label{eq:1}
\end{eqnarray}
where the first term demonstrate the sum of eigen energies $\hbar\omega_p (\hbar\omega_l)$ of SPP (LSP) modes along with $\hat{a}_{j}^\dagger(\hat{a}_{j})$ raising (lowering) operators, respectively. The second term indicates the excited state energy of QE, while ground state energy is taken as zero. $\mathcal{M}$ represents the excitation of both plasmon modes through a p-polarized light source of frequency, $\omega_o$, defined as $\sum_{j=p,l}i\hbar E_o e^{-i\omega_o t}(\hat{a}_{j}^\dagger)+h.c$ \cite{Mehmet2020,Asif2023}. The forth term defines the interaction Hamiltonian which results when QE is placed inside the hole cavity, it interacts with SPP (LSP) modes with coupling strength expressed as $\kappa_{p} (\kappa_{l})$, respectively. The last term shows the coupling of both resonant modes with interaction strength $f$. We solve the time-evolution of plasmon modes and density matrix operators $\hat{\rho}_{eg}=\ket{e}\bra{g}$, $\hat{\rho}_{ee}=\ket{e}\bra{e}$, defined in Eq.\ref{eq:1}, by using Heisenberg equations of motion, $i\hbar\partial_t\hat{a}_{j}= [\hat{\mathcal{H}}_{\rm tot},\hat{a}_{j}]$, $i\hbar\partial_t\hat{\rho}_{ge}= [\hat{\mathcal{H}}_{\rm tot},\hat{\rho}_{ge}]$ and $i\hbar\partial_t\hat{\rho}_{ee}= [\hat{\mathcal{H}}_{\rm tot},\hat{\rho}_{ee}]$. Here, we are interested merely in the field amplitudes and we disregard the quantum noise features, i.e., $\hat{a}_i=\langle \hat{a} \rangle+\delta \hat{a}=\alpha +\delta \hat{a} $ with $\langle \delta\hat{a}\rangle$=0~\cite{vitaliPRL2007optomechanical}. Hence, we replace the operators $\hat{a}_{j}$, $\hat{\rho}_{ge}$ and $\hat{\rho}_{ee}$ with complex numbers as $\alpha_{j}$, $\rho_{ge}$ and $\rho_{ee}$, respectively \cite{premaratne2017theory,Artvin_2020}. The obtained driven-dissipative dynamics are as follows;
\begin{equation}
 \dot{\alpha}_{p}= -(i\omega_p+\gamma_p)\alpha_{p}-i f \alpha_{l}-i \kappa_p \rho_{ge}+ E_o e^{-i\omega_o t}
 \label{eq:2}
\end{equation}
\begin{equation}
 \dot{\alpha}_{l}= -(i\omega_l+\gamma_l)\alpha_{l}-if \alpha_{p} -i \kappa_l \rho_{ge}+ E_o e^{-i\omega_o t}
 \label{eq:3}
\end{equation}
\begin{equation}
 \dot{\rho}_{ge}= -(i\omega_{eg}+\gamma_{eg})\rho_{ge}+ i (\kappa_p\alpha_{p} + \kappa_l\alpha_{l})y
 \label{eq:4}
\end{equation}
\begin{equation}
 \dot{\rho}_{ee}=-\gamma_{ee}\rho_{ee}+i (\kappa_p\alpha_{p}^\ast + \kappa_l\alpha_{l}^\ast)\rho_{ge}-c.c
 \label{eq:5}
\end{equation}
where y is population inversion of two-level system define as $(\rho_{ee}-\rho_{gg})$ , and $\gamma_{p,l}$, $\gamma_{eg}$ and $\gamma_{ee}$ are the decay rates of the plasmon modes and QE, respectively. 
The time-evolution of Eq.\ref{eq:2}-\ref{eq:5} is performed numerically through Runge-Kutta method in Matlab \cite{Artvin_2020,Mehmet2020} and the intensities of SPP and LSP modes are obtained as a function of excitation wavelength.\\
\begin{figure}
\includegraphics[scale=0.2]{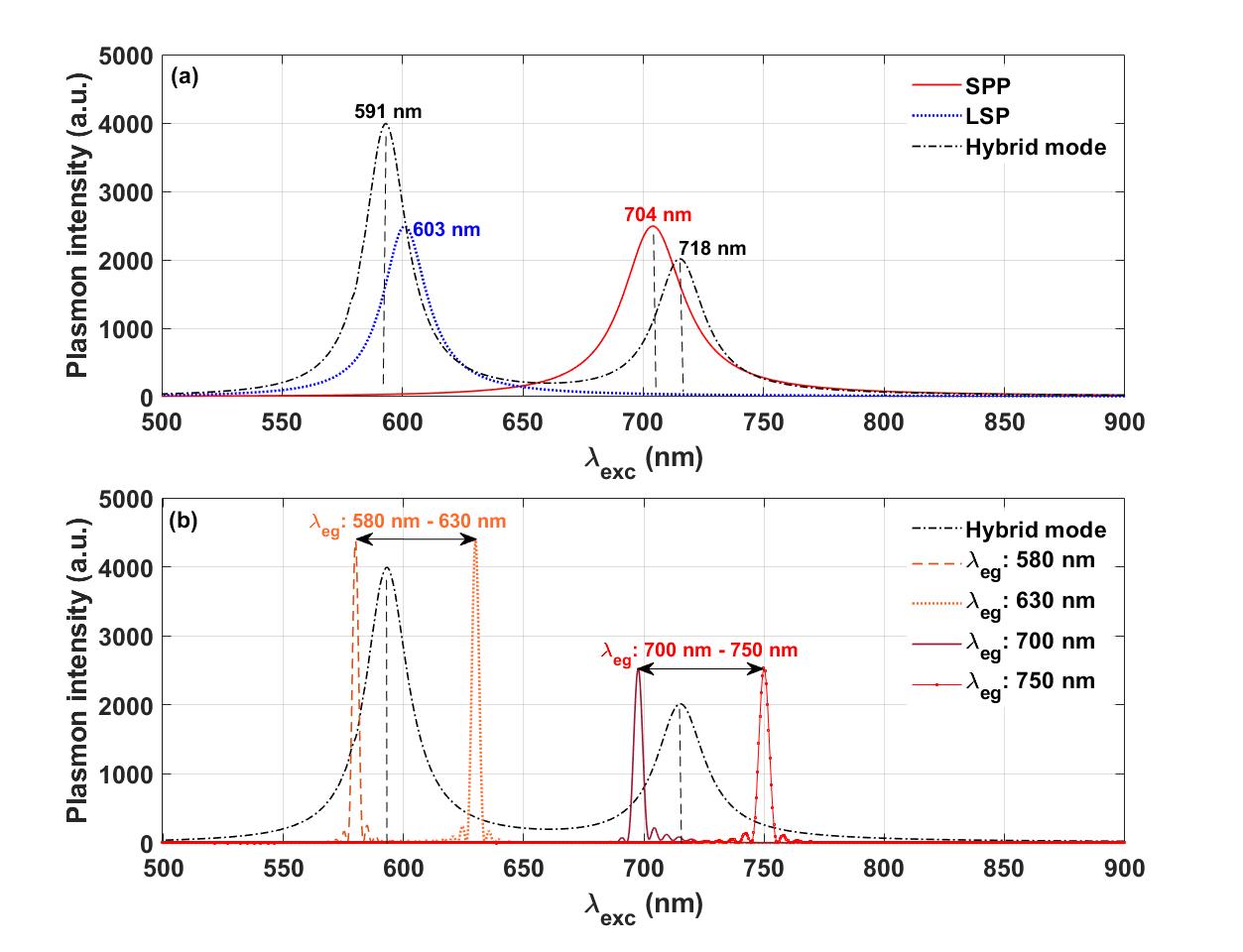}
\caption{\label{fig:2} (a) SPP and LSP field intensities as a function of excitation wavelength $\lambda_{exc}$ for coupled $(f= 0.05\omega_o)$ and uncoupled $(f =0)$ states. (b) Hybrid plasmon mode intensity (black curve) with peak resonance wavelength of SPP at 718 nm and LSP at 591 nm. The level-spacing ($\lambda_{eg}$) of QE is chosen around the resonance wavelengths of LSP and SPP modes for individual tuning.}
\end{figure}
The peak resonance wavelength of SPP and LSP modes are determined by the geometrical parameters of EOT structure. In the case of spherical hole with a diameter of 200 nm, dipole modes excited at the hole's rim exhibit a spectral wavelength of 600 nm \cite{Nehl2008}. Whereas, for a hole array with a periodicity of 400 nm, the SPP's spectral wavelength is 704 nm, drived by the plasmon dispersion relation \cite{Barnes2006}. After the resonant excitation, both plasmon modes undergo oscillations in the coupled and uncoupled states within the EOT structure. We plot the field intensities of these quantized modes with and without coupling as a function of excitation wavelength, $\lambda_{exc}$, as shown in Fig.\ref{fig:2} (a). For simplicity in calculations, all the parameters, i.e. decay rates and plasmon resonance frequencies, are scaled to the excitation frequency ($\omega_o = 2.9 \times 10^{15}$ rad/sec). In the uncoupled state $(f =0)$, LSP resonance peaks at $\lambda_{lsp}: 603$ nm, with the spectral parameters $\omega_{l}:1.08\omega_{o}$ and $\gamma_{l}:0.02\omega_{o}$. On the other hand, SPP resonance emerges at $\lambda_{spp}:704$ nm with $\omega_{p}:0.92\omega_{o}$ and $\gamma_{p}:0.02\omega_{o}$. Near the hole edge, the SPP and LSP resonances couple strongly with each other and formulate a hybrid state, owing to the interaction strength of $f: 0.05\omega_{o}$. 
In the hybrid state, the spectral wavelength of SPP mode undergoes a redshift to 718 nm, primarily resulting from the scattering interactions with the cavity boundaries. Simultaneously, at the rim of hole, intense LSP modes acquire energy from these evanescent modes \cite{Degiron2005}, leading to a blue shift in the peak spectral position at 591 nm. In section III, we will demonstrate how this modification in the spectral characteristics of oscillating modes near the cavity contributes to the far-field transmission spectrum.\\
\begin{figure}
\includegraphics[scale=0.19]{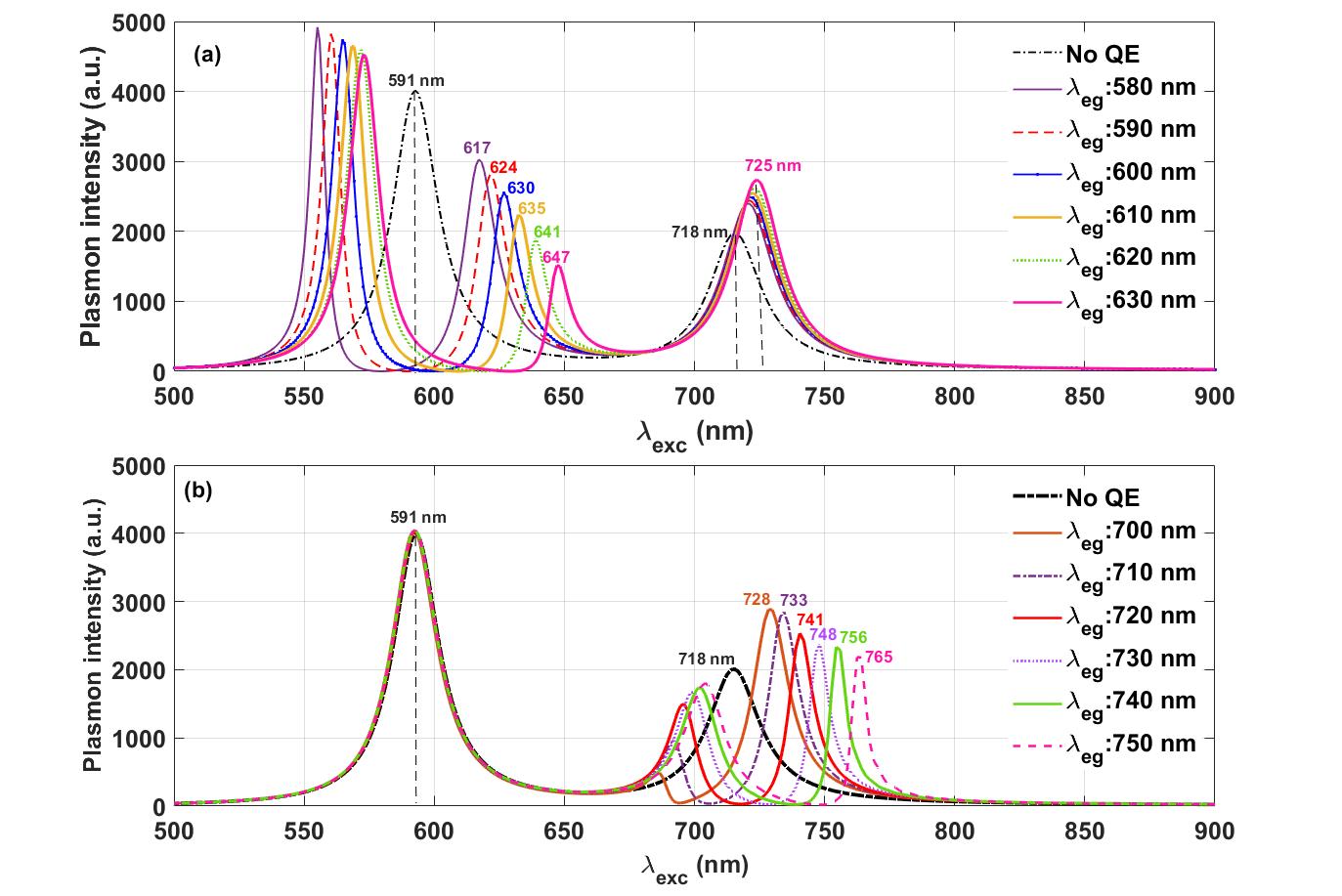}
\caption{\label{fig:3} (a) Tuning of LSP mode with QE for different transition wavelengths, ranging from $\lambda_{eg}:580$ nm to $\lambda_{eg}:630$ nm, while $\kappa_l$ is taken as $0.04\omega_o$ (b) Tuning of SPP mode coupled with QE for different transition wavelengths ranging from $\lambda_{eg}:700$ nm to $\lambda_{eg}:750$ nm and $\kappa_p$ is taken as $0.03\omega_o$.}
\end{figure}
Next, we illustrate the interaction and tuning mechanism of hybrid modes with a voltage-tunable QE in the intermediate coupling regime. For this, we place a two-level QE, characterized as a Lorentzian molecular system, inside the hole structure, as shown in Fig.\ref{fig:1}. In this configuration, QE efficiently couples with resonant modes independently, depending upon the transition frequency of the two-level system. The level-spacing of QE can be easily modified by applying an external voltage bias \cite{emre2022,Shibata2013,Yu2021}. Experimentally, one can use semiconductor materials or Stark tuned quantum dots for large spectral tuning in the visible regime \cite{Urbakh2008,Wen1995,Bawendi1997}. Moreover, TMDs and monolayer $WSe_2$ materials with large transition dipole moments also offer enhanced opto-electric tunability \cite{Gunay2020,Sie2014,darlington2023}. In Fig.\ref{fig:2} (b), we plot the peak transition wavelengths ($\lambda_{eg}$) around the resonant position of LSP and SPP modes in the hybrid state. To optimize spectral tuning, we sweep $\lambda_{eg}$ from 580 nm to 630 nm and span a difference of 10 nm between each transition wavelength. For SPP tuning, we select $\lambda_{eg}$ from $700$ nm to $750$ nm, encompassing small to large detunings with respect to mode's resonance wavelength, $\lambda_{spp}$. To see the impact of this detuning on both quantized modes, we plot plasmon intensity for different $\lambda_{eg}$, as shown in Fig.\ref{fig:3} (a,b). Fig.\ref{fig:3} (a) shows QE and LSP mode coupling. When $\lambda_{eg}$ tune to $580$ nm, $\lambda_{lsp}$ mode at 591 nm splits into two resonant peaks and redshifts to 617 nm. The dip in the curve indicates absorption at the level-spacing of QE, leading to the splitting in the LSP curve. Also, due to the pronounced field strength at the hole's edge, QE strongly couples to LSP, resulting a shift in the peaks towards longer wavelengths. Furthermore, large detuning induces a maximum redshift in the peak resonance position at 647 nm with a 10 $\%$ increase in the relative bandwidth. We observe that each tuned resonance peak emerges at a consistent 6 nm distance from the previous peak which indicates a uniform-spaced spectral shift in the resonant mode due to equidistant level-spacing of QE.\\
Similarly, Fig.\ref{fig:3} (b) demonstrate the variation in the peak spectral position of SPP mode due to modulation in the transition frequency of QE. In the absence of QE, $\lambda_{spp}$ is located at 718 nm. Upon tuning the level-spacing to 700 nm, the plasmon peak shifts to 728 nm with a $86\%$ enhancement in the peak intensity of SPP mode. This enhancement in the intensity along with the shift in resonance positions of plasmon modes, for different $\lambda_{eg}$, stems from the coherent coupling of QE with cavity resonances which enhances the interference between long and short-range SPP modes at both interfaces \cite{Berini2000}. By increasing $\lambda_{eg}$ from 720 nm to 740 nm, the peak position of SPP mode redshifts towards IR wavelengths with a decrease in the plasmon intensity. The maximum spectral shift is obtained for $\lambda_{eg}$: 750 nm, from 718 nm to 765 nm, with a total energy shift of 104 meV. To verify our analytical findings, we optimize our EOT structure by performing time-dependent numerical simulations.             
\section{Computational Approach}
\begin{figure}
\includegraphics[scale=0.4]{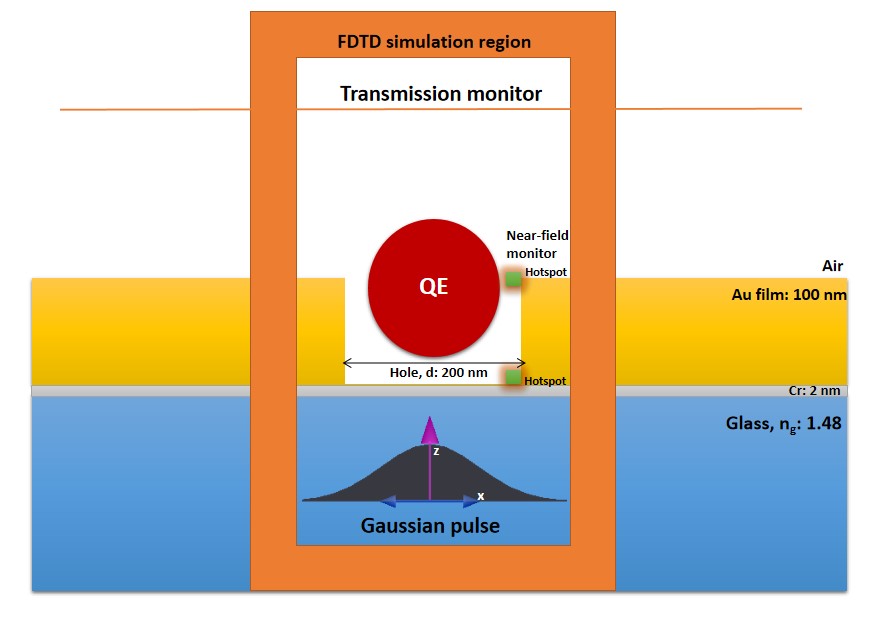}
\caption{\label{fig:4} Schematic diagram of FDTD simulation setup with a cross-sectional view of EOT structure.}
\end{figure}

\begin{figure*}
\includegraphics[scale=0.26]{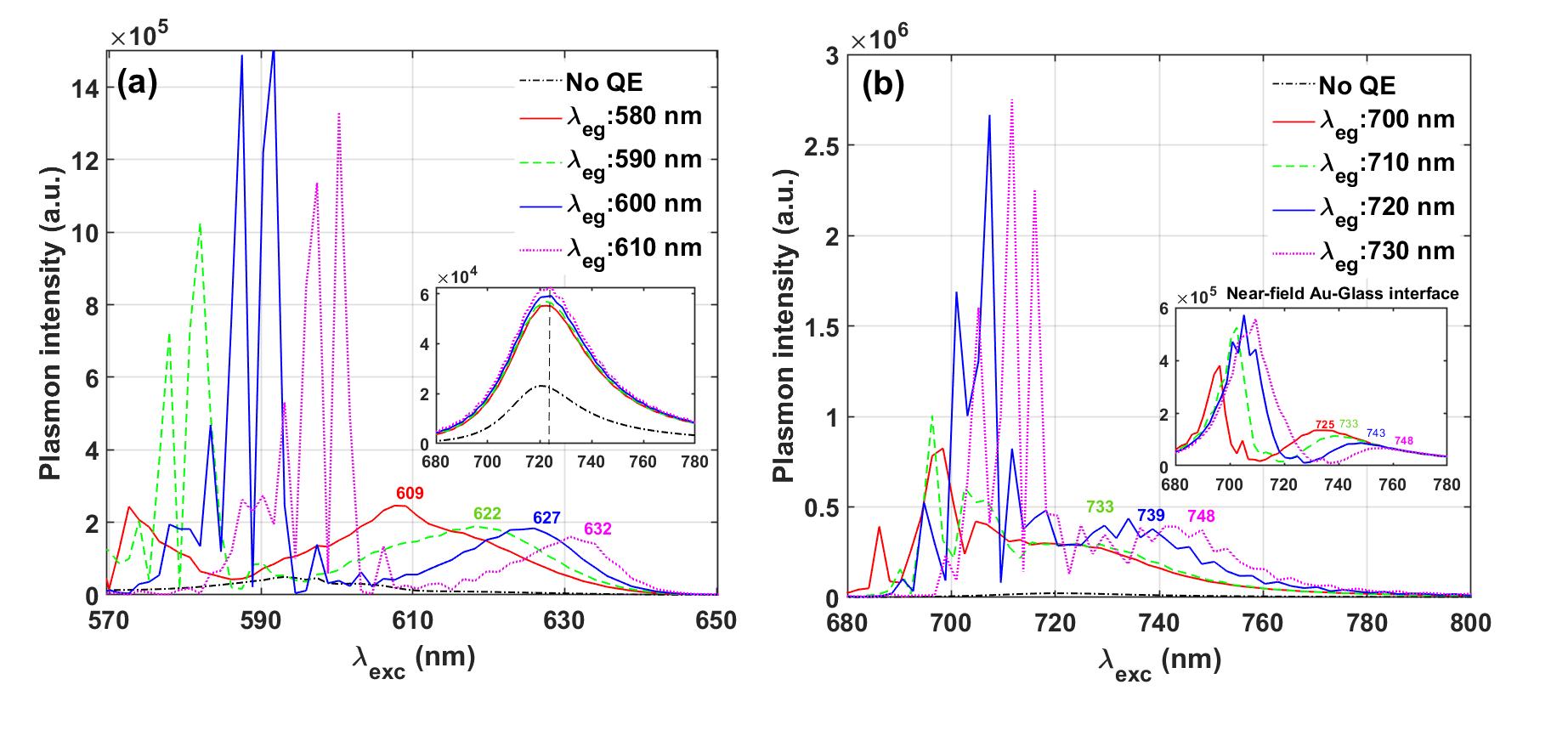}
\caption{\label{fig:5} Plasmon near-field intensity, with and without QE, as a function of excitation wavelength, measured near the hole's edge at Au-air interface. (a) Near-field modulation of LSP mode for different $\lambda_{eg}$ (b) Near-field modulation of SPP resonance for $\lambda_{eg}$. The spectral shifts in both mode resonances are indicated with wavelengths for each $\lambda_{eg}$.}
\end{figure*}

\begin{figure}
\includegraphics[scale=0.21]{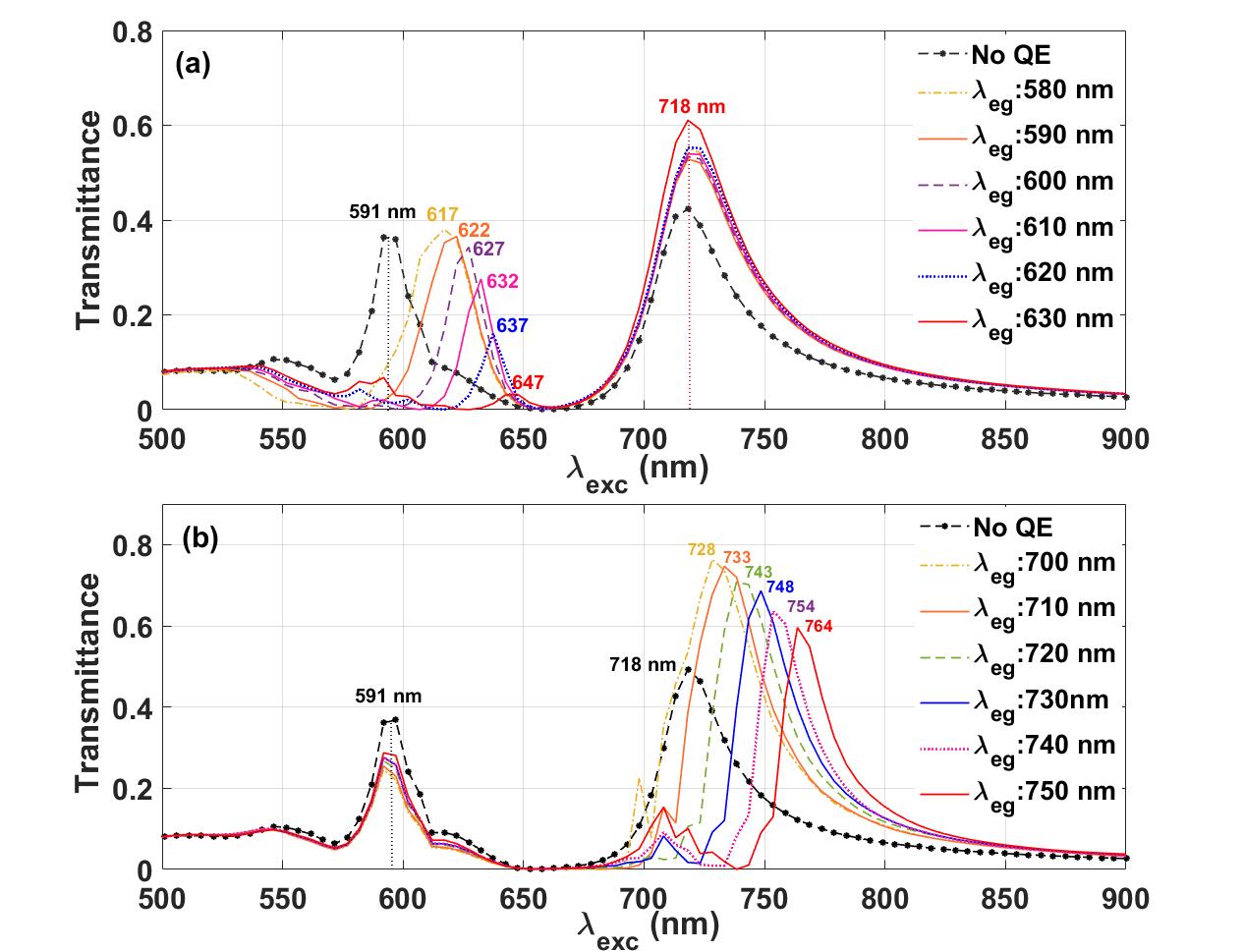}
\caption{\label{fig:6} This figure shows the normalized extraordinary optical transmission signal as a function of wavelength. After an EOT device is manufactured, its operation frequencies can be tuned by an applied voltage in this scenario. This can be performed both by utilizing the Fano resonances at the (a) LSP band and (b) SPP band of the transmission spectrum. For instance, in (a) the EOT response is very weak at 630 nm (close to operation wavelength of He-Ne lasers) without QE. When the QE resonance is electrically tuned to $\lambda_{eg}=590$ nm, however, the transmission climbs up to 0.4 (500$\%$) from 0.08. The phenomenon is even stronger in the SPP band. For $\lambda_{\rm exc}=$ 764 nm, transmission is only 0.1 for the device without QE. However, it reaches up to 0.6 when QE resonance is tuned to 750 nm.}
\end{figure}

\begin{figure}
\includegraphics[scale=0.2]{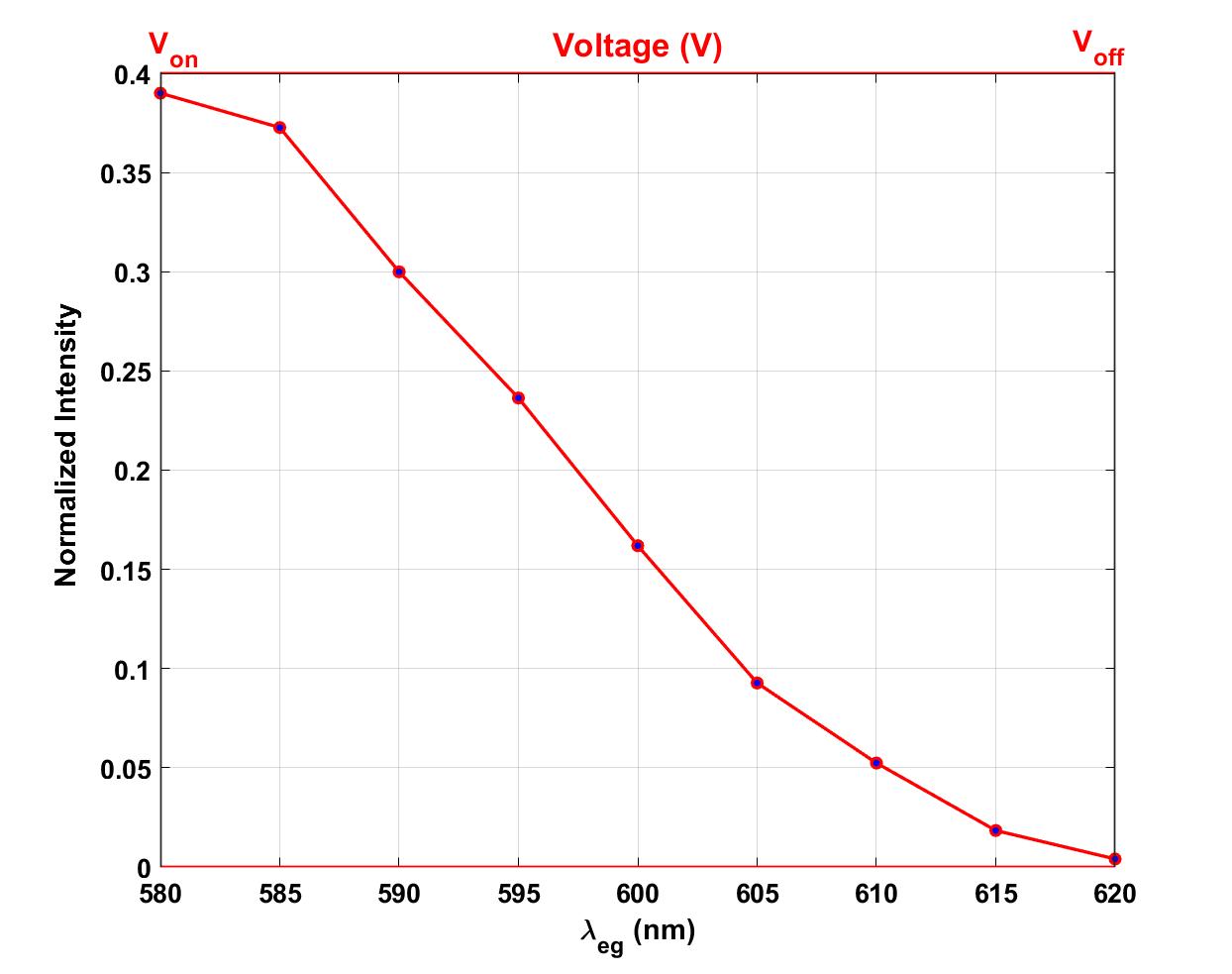}
\caption{\label{fig:7} Voltage-tunable~(programmable) EOT device operating, as an example, at $\lambda_{\rm exc}:617$ nm. Presence of the QE turns off~($3\times 10^{-4}$) the EOT signal for $\lambda_{eg}=\lambda_{\rm exc}=620$ nm and enhances it up to 0.4 as we applied bias for tuning its level-spacing to $\lambda_{eg}=580$ nm. Moreover, the EOT at $\lambda_{\rm exc}=617$ in the absence of the QE is only 0.088. More importantly, EOT signal can be continuously tuned between $\lambda_{eg}=620$ nm and $580$ nm. Such a tuning of the QE resonance can be achieved via several volts using the quantum objects which can be found in the literature, e.g., Refs.~\cite{Urbakh2008,Shibata2013}. We kindly note here that our aim is merely a proof-of-principle demonstration of the phenomenon though we use experimental values for the dielectric functions in our FDTD simulations.}
\end{figure}

In this section, we obtain the time-dependent numerical simulations results by using Finite Difference Time Domain (FDTD) method. Through FDTD method, we calculate the near-field, far-field (transmission) and time-dependent power spectra in the absence/presence of voltage tunable QE, as shown in Fig.\ref{fig:5}-\ref{fig:7}, respectively. In FDTD, we design the EOT structure in a similar way as illustrated in the analytical modeling  (Fig.\ref{fig:1}). The EOT structure consists of a 100 nm thick gold (Au) film with a circular hole of diameter 200 nm. The Au film is layered on a glass substrate with the help of 2 nm thin adhesion coating of chromium (Cr). The refractive indexes of background mediums are taken as : $n_a= 1$ for air and $n_g= 1.48$ for glass, which correspond to the top and bottom dielectric interfaces of the Au film, respectively. The frequency-dependent dielectric permittivity $\epsilon_m$ for Au is obtained from \cite{Rakic1998}. For an infinite metal-dielectric surface, a perfectly matched layer (PML) is employed in the z-direction, while periodic boundary conditions are used in the x and y directions. The FDTD simulation region encompasses a unit cell of a two-dimensional hole array situated in the xy plane, which defines the lattice constant $p$ of the array. For this particular case, the lattice constant is set to $400$ nm. The EOT structure is excited through a TM polarized Gaussian light pulse propagating in the z-direction from glass side with excitation wavelength ranging from 400 nm to 900 nm ($\lambda_{cen} : 650$ nm) and pulse duration 10 fs. Upon the incidence of the pulse, plasmon polaritons are excited at the metal-glass interface, propagating towards the hole where they interact with the dipole modes at the rim. Due to their evanescent nature, these plasmon oscillations quickly tunnel through the hole and interfere with the second-order polaritons at the metal-air interface. The collective oscillations of in-phase SPPs at both interfaces, along with LSP, enhance the surface propagation, which then scatters to the far-field as extraordinary transmission. We inspect the dynamics of coupled plasmon modes in the near and far-field region by evaluating the spectral response of the system at two different locations. For near-field spectrum, we place a near-field point monitor (green box) at 2 nm distance from the hole corner, as shown in Fig.\ref{fig:4}. While the transmission spectrum is obtained through a DFT monitor positioned 200 nm above the EOT structure on the air-side. 
To optimize EOT structure for coherent control and active tunable switching, we place a QE inside the hole cavity at 5 nm distance from the edge. For the choice of QE, we use Lorentzian dielectric function \cite{Wu2010}, $\epsilon(\omega)=1+\epsilon_{L}\omega_{eg}^2/(\omega_{eg}^2-i2\gamma_{eg}\omega_{o}-\omega_{o}^2)$ for a two-level system, where $\omega_{eg}$ and $\gamma_{eg}$ are the transition frequency and the decay rate ($10^{9}$ Hz) of QE, respectively, and $\epsilon_{L}$ is the Lorentz permittivity taken as 0.3 in the simulation.\\  
To clearly understand the increase in the peak intensity and change in the spectral position of both LSP/SPP resonances, we plot the plasmon intensity with and without QE. Fig.\ref{fig:5}(a,b) demonstrate the near-field spectra of hybrid plasmon modes for different voltage-induced emission linewidth of QE. In the absence of QE, the spectrum reveals two broad peaks at 591 nm and 718 nm, corresponding to LSP and SPP modes in the hybrid state. We couple these dominant modes individually with QE by adjusting the level-spacing. When QE is tuned around LSP resonance frequency, both peak intensity and spectral energy of hybrid mode changes in a profound way. For $\lambda_{eg}$: 580 nm, LSP mode intensity enhance upto two-orders of magnitude and resonance position shifts to 609 nm with a spectral energy shift of $62$ meV as shown in Fig.\ref{fig:5} (a). This increase in the intensity and the shift in the resonance peak is due to the exchange of oscillating energy and absorption at the QE level-spacing. For small detuning ($\lambda_{eg}-\lambda_{lsp}$), multiple LSP modes appear with maximum intensity close to resonant wavelength.
These multiple peaks results from the Rabi oscillations due to strong coupling of intense LSP modes with QE in the hole cavity. Here, we are only interested in those LSP modes which couple with propagating waves and contribute to the far-field. For this, we consider the peaks which are red-shifted from the eigen frequency. As the detuning increases, peaks shift towards longer wavelengths with a decrease in the intensity of LSP mode, while SPP mode intensity increase upto twofold at 720 nm. The inset shows the zoomed-in view of SPP peaks with enhanced intensities due to coupling with new LSP modes. When hybrid modes appear due to coupling of different resonant modes such as in waveguides and optical cavities, detuning can yield changes in their relative frequencies or energy distributions \cite{GENET2003}. As we indicated in our previous study \cite{Gunay2020}, these modulations can indeed occur even off-resonant coupling of QE and graphene plasmons. 
It can be seen that the FDTD simulation results are in clear agreement with the analytically obtained plasmon spectra (Fig.\ref{fig:3}). For SPP tuning, we switch $\lambda_{eg}$ from 700 nm to 750 nm and observe the shift in the SPP mode resonance, for clarity in the spectral position only five spectra are shown in Fig.\ref{fig:5} (b). The spectra consist of multiple resonant peaks of SPP modes in which only phase-locked interferences at the upper and lower interface enhance SPP waves before reaching to the far-field. The inset shows the spectral positions of those coherent SPP modes which excite at the Au-glass interface and then couple to the resonant modes at Au-air side. We indicate the wavelengths of these coherent modes on both interfaces with numbers (inset: Fig.\ref{fig:5} (b)). By varying $\lambda_{eg}$, the spectral shift is quite visible in both cases. For the level-spacing far off-resonant, i.e. 730 nm, mode intensity at 748 nm decreases with the increase in the linewidth at 710 nm. We analyze the impact of near-field variations on the far-field spectra by plotting the transmittance for different values of QE level-spacings, as shown in Fig.\ref{fig:6} (a,b).\\ 
The transmission spectra reveal the modulation in the spectral dynamics of SPP and LSP modes in the far-field region. As we anticipated from the near-field spectra, for different detuning frequencies, pronounced spectral shifts are observed in the EOT spectra, in the visible region. For large detuning, the peak spectral positions shift towards longer wavelengths, due to change in the energy and relative phase of plasmon resonances at both interface. In Fig.\ref{fig:6} (a), $\lambda_{lsp}: 591$ nm, after coupling to $\lambda_{eg}: 580$ nm, shifts to 617 nm with a total energy shift of 84 meV. On the other hand, at $\lambda_{spp}: 718$ nm, the transmission efficiency is enhanced upto $61\%$ as compared to the system with no QE. However, the shift in LSP peak results from the perturbation induced by QE emission linewidth which changes the interference correlation between multiple LSP modes at the rim. For every tuned level-spacing (Fig, \ref{fig:6} (a)), the difference between each LSP shift is around 5 to 10 nm, while maximum spectral shift produce at 647 nm with an energy of 181 meV. In Fig.\ref{fig:6} (b), we observe a 55 $\%$ enhancement in the transmission intensity at 728 nm peak for $\lambda_{eg}$: 700 nm, and the redshift in the energy is around 23 meV. The peaks are uniformly shifted to higher wavelengths with a 10 nm distance in between. As $\lambda_{eg}$ tune to far off-resonance level-spacing, 750 nm, the transmission peak acquire maximum energy shift upto 104 meV. The change in the energy density of polaritonic states results from the modification in the local-density of states (LDOS) while interfering with multiple resonant and off-resonant modes in the structure. Also the strong oscillation of QE influence the propagating direction of polaritons at both surfaces, which originates new SPP peaks in EOT spectra.
\subsection*{(i) Off-resonant operation of EOT device}
Photonic devices are generally operated at their resonance frequency to achieve high throughput. However, this limits the device capability when it is desired to operate at a different frequency. Consequently, time-consuming processes such as design, production, and testing stages are required. In contrast, our proposal eliminates these limitations. It's noteworthy that this device can potentially function at alternative excitation wavelengths as explained as follows, provided that an appropriate QE is employed. The coupling between QE and plasmon mode manifests coherent correlation of quantized modes which induces a shift in LSP (or SPP), independently. More explicitly, the EOT intensity, e.g., at $\lambda_{exc}=632$ nm is 0.04 without the QE. However, when we employe a QE in our system ($\lambda_{eg}=610$ nm) the transmission reaches 0.28 at the smae wavelength $\lambda_{exc}=632$ nm (see Fig. \ref{fig:6} (a)). In addition, one can observe this phenomena at the other operation frequencies via applying bias-voltage to QE. To the best of our knowledge, the development of a voltage-tunable EOT device with such capability has not been achieved yet. Here, we demonstrated in principle that the suggested EOT device operates specifically around the wavelengths of localized surface plasmon resonance. On the other hand, the same mechanism can also be achieved around SPP wavelengths ($\sim$730 nm). For example, we assume that we placed a QE into nanohole ($\lambda_{eg}=750$ nm) and we measure the EOT signal at $\lambda_{exc}=764$ nm. As it can be seen from Fig. \ref{fig:6} (b), the EOT signal is increased from 0.12 (without QE) to 0.60 (with QE).\\

\subsection*{(ii) Electrically-programmable EOT device}
Taking the spectral modulation of EOT signal into account, our proposed device can be used as quantum switch in a simple structure. A fixed level-spacing of QE yields a spectral shift in plasmon modes with maximum intensity. If we apply voltage to QE, the tuned transition wavelength modulates the EOT signal intensity in the far-field from minimum (off-state) to maximum (on-state). In this way, a dynamical QE system with various spectral response can be used for the voltage tunable quantum EOT switch. To illustrate this, we plot intensity of EOT signal measured at $\lambda_{exc}=617$ nm (see Fig.\ref{fig:7}) as a function of tuned transition wavelength ($\lambda_{eg}$) of QE after employing bias-voltage. We assume that the initial level spacing of the QE is 620 nm, and it gradually shifts to the left as several volts are applied. In addition, we estimated the amount of shift in the level spacing of QE in response to the applied voltage using \cite{Shibata2013,Urbakh2008}. The placing of QE shifts the resonances of LSP or SPP modes to wavelengths where the EOT device originally exhibited no significant response (see Fig. \ref{fig:6} (a) and (b)). Therefore, the EOT device can be operated off-resonantly (as if it is on-resonance excitation) by exciting it at a specific wavelength, denoted as $\lambda_{exc}=617$ nm, while concurrently measuring the transmission intensity at that wavelength. This evaluation transpires alongside the application of an external bias voltage, which modulates the transmission, a scenario depicted in Fig.\ref{fig:7}. Remarkably, the transmission reaches its maximum at $V_{on}$.\\
Along with high oscillation strength, the lifetime of excited state of QE is $\sim$ ns, which is quite long compared to ultrafast plasmon oscillations ($\sim$ 10 fs) \cite{Kirakosyan2016}. The difference in the decay rate characteristics of both oscillators helps to change the temporal dynamics of fast decaying plasmon field \cite{Yildiz2020}. The lifetime extension of both linear and nonlinear plasmon modes, in the vicinity of QE, have been discussed in \cite{Yildiz2020,Ovali2021,Asif2022} with detailed analysis.
\begin{figure}
\includegraphics[scale=0.2]{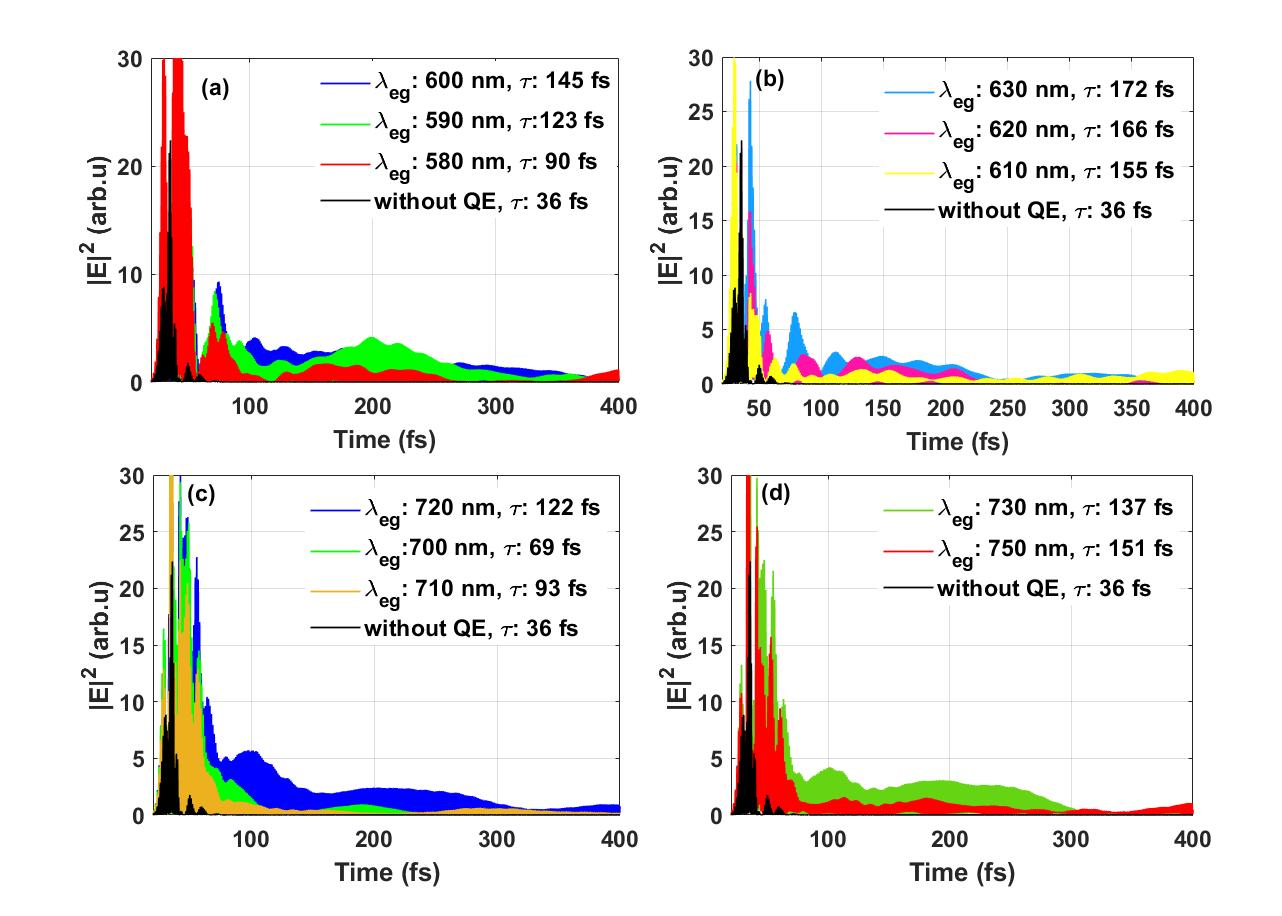}
\caption{\label{fig:8} Near-field power spectra of plasmon field intensity for tuned resonant wavelengths in the presence/absence of QE. (a,b) Plasmon oscillations for different $\lambda_{eg}$ and calculated average lifetime at each tuned LSP wavelength. (c,d) Plasmon intensity measured for different $\lambda_{eg}$ with average oscillation time at tuned SPP wavelengths.}
\end{figure}
However, in this study, we anticipate lifetime enhancement phenomenon only at the tuned spectral positions of the plasmon resonances. In this way, along with fine spectral tuning of transmitted signal, we can also modulate the coherence time of EOT device simultaneously on a single chip. This process not only improve device efficiency but also ensure signal stability for long time. To optimize the change in the average lifetime of plasmon mode, we plot plasmon field intensity as a function of time. Fig.\ref{fig:8} (a-d) illustrates the modulation in the field oscillation of (a,b) LSP and (c,d) SPP modes for different tuned resonant wavelengths. When there is no QE in the system, plasmon modes in the hybrid state decays quickly with an average lifetime ($\tau$) of 36 fs. However, with QE in the system, the tunable characteristics modulate the spectral phases of both modes significantly, giving rise to enhanced temporal efficiencies of the plasmonic system. We calculate the average lifetime of power spectra for different resonance position from, $\tau=\int{t\:I(t)} dt/\int{I(t)} dt$, where I(t) is the plasmon intensity and t is time \cite{Asif2023}. In Fig.\ref{fig:8} (a), plasmon oscillations shows a 50 $\%$ enhancement in the lifetime at peak $\lambda_{lsp}:$ 617 nm (for $\lambda_{eg}:$ 580 nm tuning) compared to the case with no QE in the system.  This is merely due to increase in the mode's intensity and temporal bandwidth in the plasmon-emitter coupled state. For large detuning, the emission linewidth $\lambda_{eg}:$ 630 nm induces the spectral shift located at $\lambda_{lsp}: 647$ nm with a maximum lifetime enhancement upto 172 fs (Fig.\ref{fig:8} (b)). In the case of SPP, a uniform shift in the energy along with the average oscillation time can be seen in Fig.\ref{fig:8} (c,d). With a maximum spectral shift and large oscillation amplitude, the transmission mode peaks at 764 nm (for $\lambda_{eg}:$ 750 nm tuning) has the longest oscillation time of 151 fs. The voltage-induced variation in the frequency and linewidth of QE not only tune the spectral bandwidth of the EOT light but also modulate the coherence time at desired photon energy.    
\section{Conclusion} 
We demonstrated active control of extraordinary transmission of light through the voltage-tunable QE system. The spectral and temporal modulation of transmitted light is optimized by coupling hybrid plasmon modes with coherent oscillating two-level QE. By varying the level-spacing of QE through external bias voltage, we obtained spectral and temporal modulation in the SPP and LSP resonances. Our results clearly demonstrate that under the intermediate coupling the resonance frequency of hybrid modes redshifts to longer wavelengths with a maximum spectral shift upto 181 meV.  Also, the coherent coupling of plasmon modes at both interfaces, in the presence of QE, increase the oscillation efficiency for longer time which in turns enhance the relaxation time of hybrid modes upto 172 fs for $\lambda_{lsp}: 647$ nm and 151 fs for $\lambda_{spp}: 764$ nm. This long oscillation time of hybrid plasmon-emitter system incorporates in the average lifetime enhancement of transmitted light at the shifted wavelengths only.
Our proposed EOT device with tuned spectral and power functionality can be integrated with light display or utilized as Quantum switch in photonic integrated circuits (PIC) and quantum information processing, respectively. In addition, the high temporal and spectral sensitivity of such EOT structures makes them suitable for use in bio-chemical sensing, imaging, solar energy harvesting and spectroscopy of individual molecular systems. 
 
\begin{acknowledgments}
R.S., M.E.T., and H.A. acknowledge support from TUBITAK-3501 grant no. 121F030 and TUBITAK-1001 grant no. 123F156. MET acknowledge support from TUBITAK-1001 grant no 121F141.
\end{acknowledgments}


\providecommand{\noopsort}[1]{}\providecommand{\singleletter}[1]{#1}%

\end{document}